# A Total Solar Eclipse Earth-Based Mission: Multi-wavelength Observations from Land, Sea and Air to Probe the Critical middle Corona

*A White Paper for the 2024 - 2033 Solar, Space Physics and Heliophysics (SSPH) Decadal Study*

Shadia R Habbal, Institute for Astronomy, University of Hawaii, HI
Benjamin Boe, Institute for Astronomy, University of Hawaii, HI
Colby Haggerty, Institute for Astronomy, University of Hawaii, HI
Adalbert Ding, Technical University Berlin and Institute for Technische Physik, Germany

## SYNOPSIS

Space exploration in the late 1960s enabled imaging of the solar corona both in projection on the disk and off the limb in the extreme ultraviolet and x-rays for the first time. The success of these first observatories led to the discovery of the intricate role of magnetic fields in shaping coronal structures. They also drove technological and scientific advances over more than half a century, resulting in exquisite high temporal and spatial resolution coronal emission data. However, the inherent limited spatial extent of emission at these wavelengths precludes exploration of the intricate changes in the quiescent and dynamic plasmas and magnetic fields between 1.5 and 6 solar radii ($R_s$), which we refer to as the middle corona. On the other hand, significant technological advances in space-based white light coronagraphic imaging have currently reached an almost unlimited field of view; yet, they lack the diagnostic tools for probing key coronal plasma parameters. Indeed, there has been an unfortunate gap in coronal emission line observations from space in the visible and near IR (V+NIR). Their distinct scientific advantage stems from the dominance of radiative excitation in their formation, whereby their emission can be detected out to several solar radii above the limb. V+NIR emission lines can thus yield the only inferences of the physical properties of the coronal plasma, such as species temperatures, densities, elemental abundances, and speeds along and perpendicular to the line of sight in this critical spatial span. These diagnostics have been demonstrated with decades of unsurpassed high-resolution imaging and spectroscopic observations during total solar eclipses.

**This white paper calls for dedicated funding for a Total Solar Eclipse Earth-Based Mission for ground, airborne and seaborne observations of the corona during totality for the next decade starting in 2024. The proposed Mission capitalizes on the unique diagnostic potential offered by the V+NIR coronal emission lines for the inference of key plasma parameters over a distance range of at least 5 $R_s$ from the solar surface. This critical coronal space is currently missing from existing and to-be launched coronagraphic instrumentation in the proposed time frame.** Multi-site observing platforms for each eclipse would further capture the temporal variability of coronal plasmas over a time span of at least 1 hour, with a temporal resolution of a fraction of a minute. Furthermore, this Mission offers unsurpassed opportunities for the exploration of new technologies for future implementation with coronagraphs. This Mission has a unique significant broader impact for outreach opportunities to engage the public and the younger generations in heliospheric science from an awe-inspiring cosmic event.

# 1. Science Goals: Impetus for a Total Solar Eclipse Earth-Based Mission

*1.1 The Unique Scientific Potential of Total Solar Eclipse Observations*

Despite the exceptional spatial and temporal resolution of EUV and X-ray observations, they are inherently excluded from probing the physical properties of the coronal plasmas and fields in a spatially uninterrupted field of view extending past 1.5 $R_s$ (Fig. 1A). This is a critical space where the coronal plasmas and magnetic fields evolve most rapidly. The connectivity between complex coronal structures detected in space-based white light coronagraph images and their sources at the Sun, in particular at solar maximum, is thus missing from these observations, as well as the identification of the sources of the solar wind and dynamic events at the Sun (Figs. 1 A and 1B). At present, only total solar eclipse (TSE) images in white light, visible and near-infrared (V+NIR) emission lines provide the continuous coverage necessary to fill this critical gap. Unfortunately, their diagnostic potential is missing in present and soon to-be launched platforms.

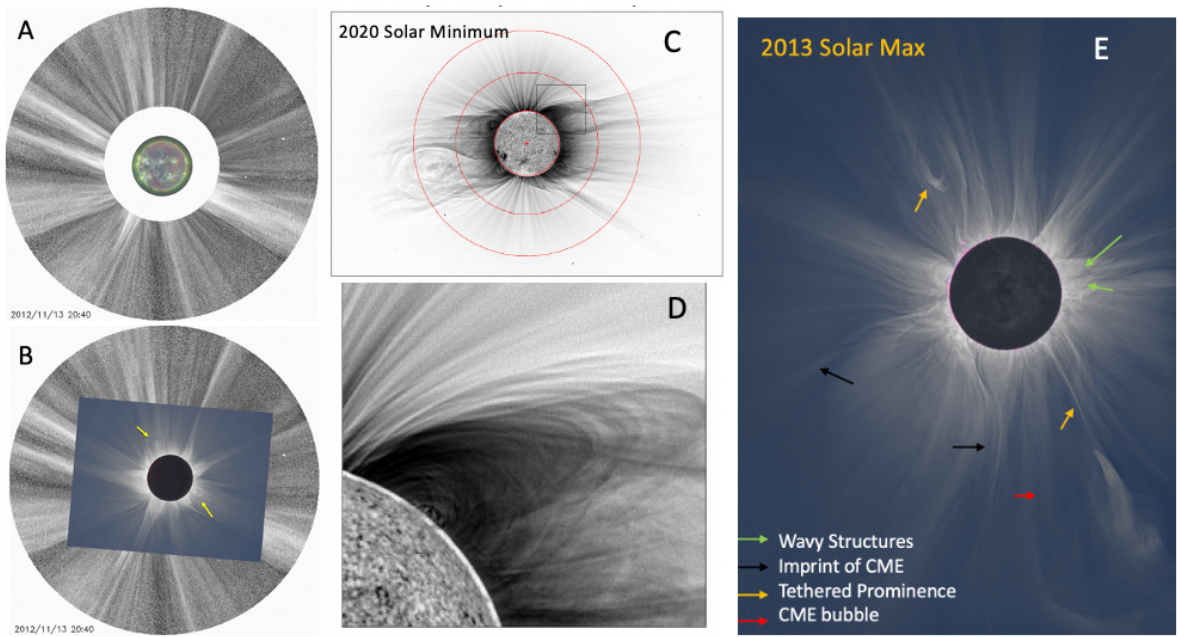

*Figure 1: (A) LASCO/C2 coronagraph image from the 2012/11/13 total solar eclipse, with corresponding AIA composite image revealing a critical data gap, only filled by a white light eclipse image (overlay in B) to establish the connectivity between coronal structures in C2 and their sources in the corona (e.g. two emerging CMEs indicated by the yellow arrows). (C) 2020 eclipse white light image at solar minimum with the inner edge of the LASCO/C2 and C3 occulters given by the red rings at 2.2. and 3.5 $R_s$. (D) Fine details of coronal structures arches overlying a prominence from box in C. (E) Plethora of dynamic events: waves, tethered erupting prominences and CMEs at solar maximum (see annotated colored arrows).*

Indeed, high resolution white light eclipse images are unique for capturing the instantaneous status of dynamic coronal structures such as prominence eruptions, coronal mass ejections (CMEs) and their unmistakable tethers to the Sun (Figs. 1B, 1C) (Habbal et al. 2014; Alzate et al. 2017). They also capture prominences within the context of large-scale overarching coronal structures, their connectivity to them (Fig. 1D, Figs 2A-B), as well as their association with different turbulent structures (Fig. 2D) (Druckmüller et al. 2014), CMEs (Fig. 2E), corkscrew-type structures following the passage of a CME (Fig. 2F), and identification of a double-bubble CME (Boe et al. 2021). Fig. 2D shows how atypical coronal structures, akin to plasma instabilities, identified in eclipse images (Druckmüller et al. 2014), appear in WISPR on Parker Solar Probe.

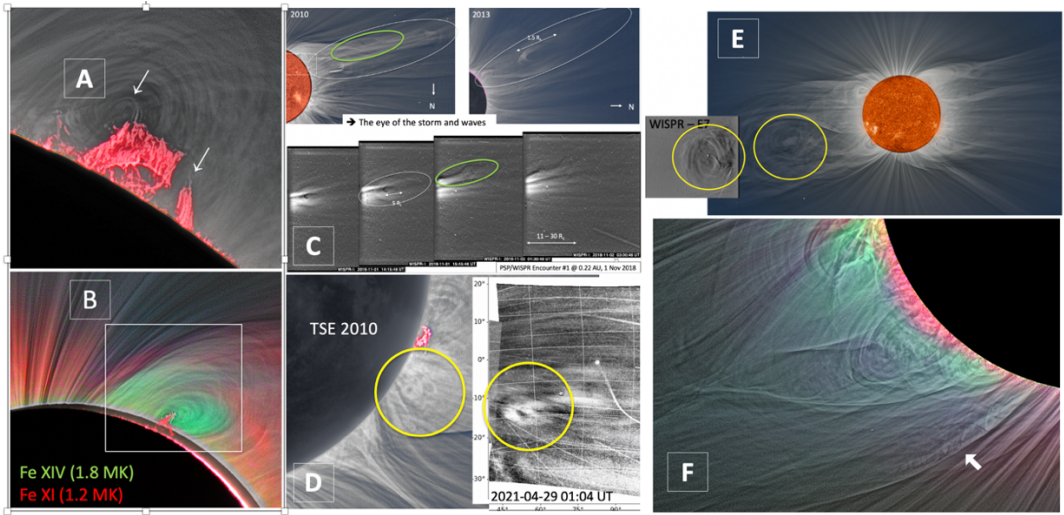

*Figure 2: (A) 2017: Connectivity between a prominence and overlying arch-like structure, identified as hot shrouds in Fe XIV emission (B). (C) Different stages of prominence eruption from 2010 & 2013, with identical structures in WISPR (bottom). (D) 2010: Emergence of turbulence and plasma instabilities in the vicinity of prominences, with similar structures in WISPR (right). (E) 2019: Complex CME filamentary structure, and comparison with WISPR (inset). (F) TSE 2017: Screw-type structure at edge of a CME.*

Furthermore, the suite of Fe emission lines in the V+NIR (Fig. 3B), first discovered during the 1869 eclipse (Young 1872), underscore their unique scientific yield (see white light and color composite of white light, Fe XI and Fe XIV in Fig. 3A), for mapping the distribution of the electron temperature, $T_e$ (Habbal et al. 2010a, 2021), and the inference of $T_e$ (Fig. 3C). White light and multi V+NIR imaging yield the distribution of the electron density associated with the magnetic topology of coronal structures; see Boe et al. 2020) and corresponding emission from Fe XI 789.2 nm at 1.2 MK and Fe XIV 530.3 nm at 1.8 MK (Boe et al. 2018). At present, multiwavelength TSE images (Fig. 3A) are the only observations to yield ionic abundances, and the empirical inference of the ion freeze-in distances (Fig. 4; Habbal et al. 2007, Boe et al. 2018). This distance is determined by the distribution of $T_e$ and density in the corona, thus reflecting the heating processes responsible for this distribution [See Boe et al. White Paper].

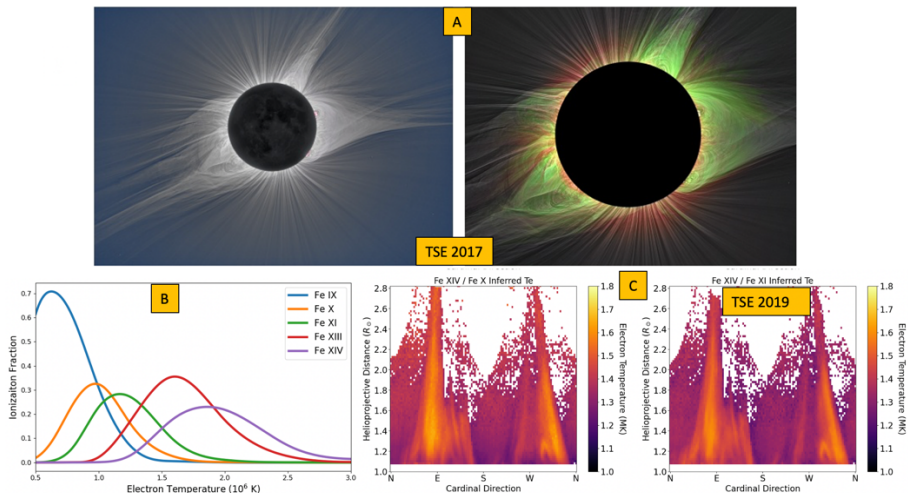

*Figure 3: (A) White light eclipse image (left) and corresponding composite image $T_e$ map with Fe XIV 530.3 nm (1.8 MK) and Fe XI 789.2 nm (1.2 MK) emission (right). (B) Ionization fraction of the Fe coronal emission line sequence versus $T_e$. (C) Inferred $T_e$ from Fe X, Fe XI, Fe XIV intensity ratios. (From Boe et al. 2018, 2022).*

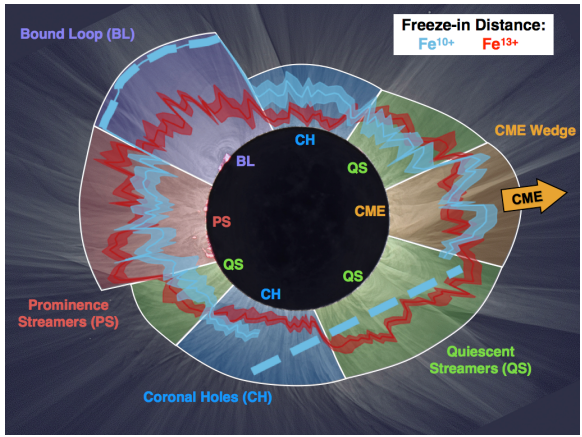

Figure 4: Overlay of the white light 2015 total solar eclipse image of the freeze-in distance inferred from Fe 10+ (blue) and Fe 13+ (red) and continuum 2015 [Boe et al. 2018].

Imaging spectroscopy during total solar eclipses also continues to yield unique data. In particular, the 2015 total solar eclipse yielded novel insights into the fate of prominence material following a prominence-CME eruption (Ding and Habbal 2017) (see Fig. 5). Doppler shifted emission from neutrals, low-ionized elements as well as Fe XIV, was detected in the corona, with speeds of 100 to 1500 km/s 1-2 $R_s$. These observations confirmed that in-situ measured neutrals and low-ionized ions originate from CME-associated prominence eruptions. Koutchmy et al. (2019) ascertained the presence of Ar emission in the corona. Samra et al.'s (2018, 2019, 2022a,b) spectra in the NIR led to the discovery of new coronal lines, ideally suited for coronal magnetometry.

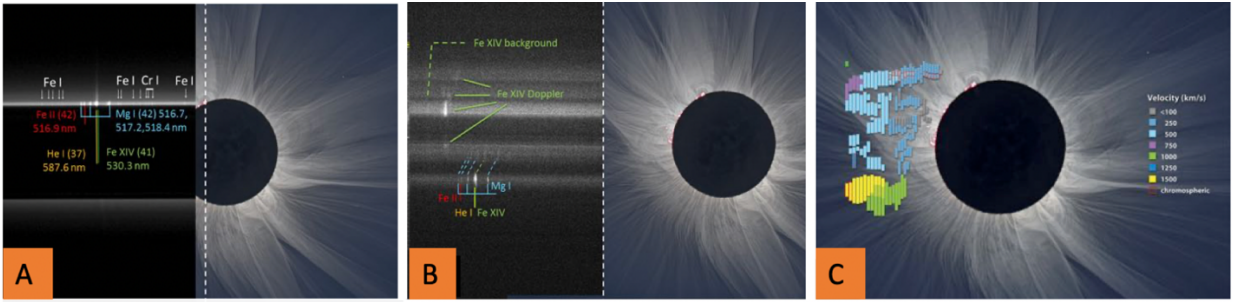

Figure 5: Spectral observations from 2015 with different slit positions (A and B) relative to coronal structures. (C) Mapping of the Doppler shifted coronal and prominence material (Ding & Habbal 2017).

White light and multi-wavelength eclipse observations spanning more than a solar cycle, together with in-situ Fe charge state measurements from ACE spanning the same time period, recently led to the first time identification of the sources of the slow, intermediate and fast solar wind at the Sun (Habbal et al. 2021). The dominance of $Fe^{10+}$ in-situ charge state was directly connected to the ubiquitous Fe XI emission (from $Fe^{10+}$) in open field lines throughout the corona (see Fig. 6). The in-situ high charge states or $Fe^{13+}$ and higher were associated with hot emission from CMEs in the corona, such as by Fe XIV emission.

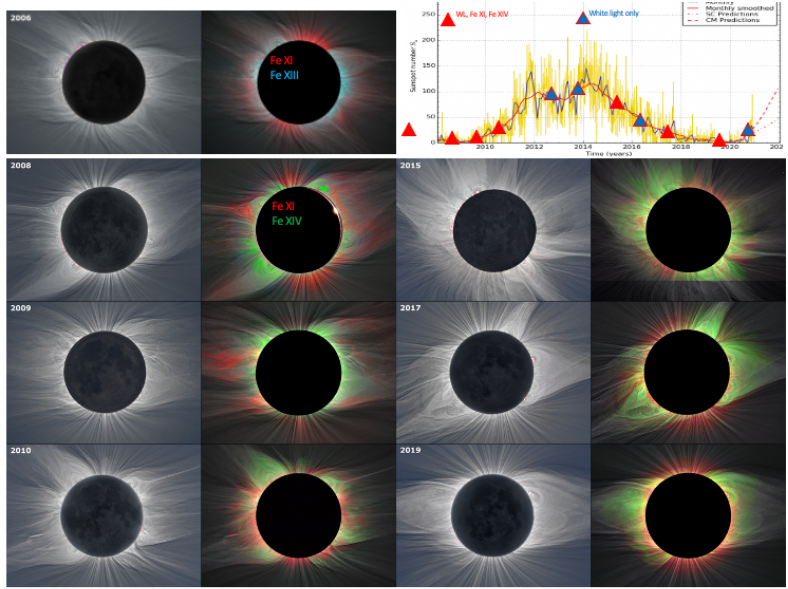
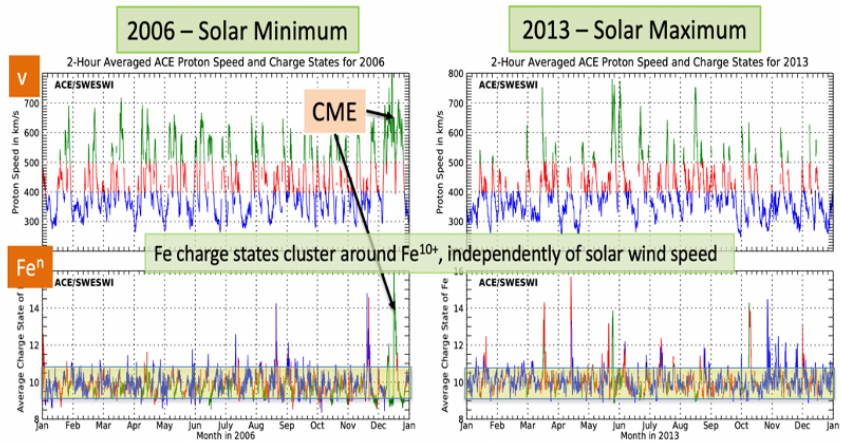

*Figure 6: Top: Columns of white light and FeXI/Fe XIV composite eclipse images acquired between 2006 and 2019. The plot at the top right is the sunspot number during that time period. Red triangles correspond to dates with Fe XI/Fe XIV and white light observations. Blue triangles are for missing Fe XI/Fe XIV observations due to bad weather. Bottom: Fe charge state measurements from ACE from 2006 (solar min) and 2013 (solar max) showing a ubiquitous band of in-situ charge states centered on $Fe^{10}$, independently of solar cycle. [Habbal et al. 2021].*

*1.2. Science Goals*

These novel insights into the plasma properties of the quiescent and dynamic status of the corona and the sources of the solar wind, gleaned from the unsurpassed spatial and spectral resolution of total solar eclipse observations in the V+NIR, over an uninterrupted spatial extent of $1 - 5 \; R_s$, are currently absent from existing ground and space-based coronagraphs. They are the impetus for the science and implementation goals of this **Total Solar Eclipse Earth-based Mission,** namely:

- Acquire high resolution white light and multi-wavelength V+NIR total solar eclipse observations over 5-10 $R_s$ above the solar surface, for a duration comparable to a solar cycle, starting in 2024.
- Maximize the temporal span of eclipse observations along the path of totality, to overcome weather inclement and further exploit temporal changes in, e.g., $T_e$ and composition, initiation of waves and instabilities by prominence eruption and CME passage (see Boe et al. 2020a), over tens to hundreds of minutes.
- Develop state of the art instrumentation for imaging and spectroscopic V+NIR observations which capitalizes on technological advances.
- Develop and test next generation technology for space-based coronagraphic V+NIR imaging and spectroscopic imaging for continual observations of the corona.

2. **The Proposed Mission**

The proposed Mission calls for a suite of ground-based, airborne and seaborne platforms, to acquire observations throughout the era of new spacecraft cruising through unchartered territory, such as the Parker Solar Probe (PSP), Solar Orbiter (SO), and PUNCH. Not only will it offer a critical element for the exploration of the physical properties of the corona and the sources of the solar wind streams, it will also charter new scientific frontiers in the exploration of the solar corona, as demonstrated by past eclipse discoveries.

*2.1. Proposed Schedule of Total Solar Eclipse Observations and Observing Platforms*

Total solar eclipse observations to be acquired from a suite of ground-based, airborne and seaborne platforms, covering the next 8 total solar eclipses from 2024 to 2035, thus spanning the duration of a solar cycle, are recommended for this Total Solar Eclipse Earth-Based Mission. The compilation in Fig. 7 from https://time.com/4897581/total-solar-eclipse-years-next/ gives durations of 2 - 6.4 min, with a 50% occurrence over land, and the rest over oceans. The proposed suite of platforms will maximize data acquisition to overcome weather inclement (which is always a lurking threat with any ground-based observations), and temporal coverage spanning the duration of totality across Earth. Multiple ground-based observing sites, remain the optimal choice from the point of view of accessibility and stability. They can be selected to provide an almost continuous coverage along the path of totality to detect changes in the corona on time scales of seconds to minutes; this is particularly relevant in the advent of a CME (see Boe et al. 2020a, 2021). For example, the 2024 TSE will take ~ 5 hours across the whole path, which is equivalent to 8 $R_s$ travel span across the corona for an average CME speed of 300 km/s.

Airborne platforms, such as kites, drones and high-flying aircraft, offer opportunities to overcome the threat of clouds, and extend observations over oceans. While more challenging, airborne observations have the added advantage of reduced airmasses, thus increasing the signal to noise ratio when compared to ground-based observations.

*2.2. Basic Eclipse Observations and Types of Instrumentation*

To maximize the science yield of the proposed eclipse observations offered by the rich V+NIR coronal emission, a main consideration for instrumentation is the limited duration of totality, and the size of the equipment (namely transportation considerations given the sporadic nature of eclipse occurrences across the world). Examples of instrument types with proven heritage are provided.

- White light and continuum imaging, with and without polarization with an angular resolution of 1" < 1.5 $R_s$, and 5" up to 10 $R_s$. These observations can be achieved with high-end commercially available cameras and lenses. A suite of different focal lengths and exposure times ensures the coverage of a significant fraction of the coronal space.
- Multi-wavelength V+ NIR imaging, with and without polarization. A major requirement for V+NIR imaging is the subtraction of continuum emission underlying the emission lines. This can be achieved with filters tuned to 3.0 – 5.0 nm from centerline, operated simultaneously with filters centered on distinct emission lines. Choices of different focal length lenses (e.g. from 100 to 500 mm*)* yield different coronal spatial coverages which can be combined post-observation. Technologies are currently available for adding polarization capabilities, such as the use of *polarcams* with embedded polarizers on the pixels.

- Imaging spectroscopy continues to be a powerful technique for exploiting the coronal spectrum and discovery of spectral lines. These measurements are the only ones to yield Doppler widths, essential for the inference of species temperatures, and Doppler shifts when present, to yield material motions along the line of sight. Different spectrometer designs have been developed by different groups (e.g. Koutchmy et al. (2019), Pasachoff et al. (2019), Samra et al. (2018 – 2022b)). Imaging spectroscopy can be achieved by enabling the slit of the spectrometer to spatially scan the corona at set time intervals during totality. For example, a partially multiplexed design maximizes the yield of a spectrometer over the V+NIR spectral range and enables the detection of coronal as well as chromospheric emission (see Ding and Habbal 2017). Other approaches have involved positioning the slit at select locations in the corona (e.g. Samra et al. 2018, 2019, 2022a,b). Regardless of the choice of spectrometer design, a critical factor is to ensure the recording of the position of the spectrometer slit within the context of the corona during totality, often referred to as a slit-jaw or monitor camera.

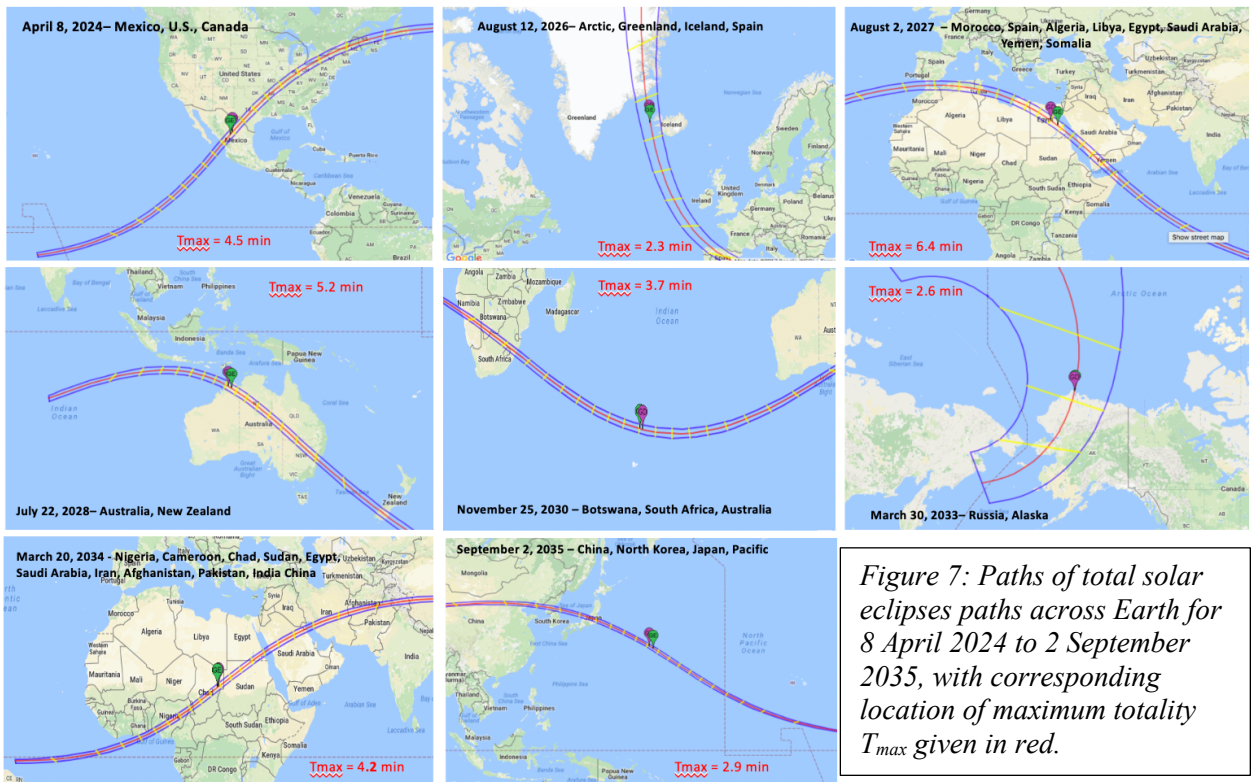

Figure 7: Paths of total solar eclipses paths across Earth for 8 April 2024 to 2 September 2035, with corresponding location of maximum totality $T_{max}$ given in red.

2.3. Proposed multi-site platforms:
- Ground-based
- Seaborne
- Airborne platforms: including drones, kites, balloons and high altitude research aircraft.

Kites and drones can be launched from multiple sites from land and from vessels at multiple sites, and can carry payloads of a few 10's of kg. Specially designed high altitude research aircraft (such as the NASA WB-57) are another option, which further extend the duration of eclipse observations by a few minutes. They also extend the wavelength range to the ultraviolet down to 270 nm thus enabling access to an abundance of untapped diagnostic spectral lines.

*2.4. Innovative technological advances*

A unique advantage of eclipse instrumentation is access to innovative technological advances on a rather fast time schedule, unlike space-based instrumentation which become locked at launch. These opportunities also enable innovative compact instrumentation designs which are ideal for future space-based implementations. As such they can also be considered as heritage experimentation for space, albeit the latter requiring the use of a coronagraph.

### 3. Broader Impact and Outreach

The unique science gleaned from total solar eclipse observations as well as the witnessing of a cosmic event of unsurpassed beauty can readily instill inspiration in the younger generation. These outstanding astronomical 'field expeditions' in astronomy can provide unsurpassed opportunities for undergraduates, graduate students and postdoctoral fellows to get involved in the construction, deployment of novel instrumentation for the eclipses, data collection and analysis for a wide range of projects. The proposed Mission underscores the necessity to ensure the continuity of total solar eclipse observations independently of coronagraphic observations as they remain unique opportunities for scientific explorations of the corona.